\documentclass[aps,pra,showpacs,twocolumn]{revtex4-1}
\usepackage{amssymb}
\usepackage{amsmath}
\usepackage{graphicx}
\usepackage{epsfig}

\begin{document}

\title{Geometric phase and phase diagram for non-Hermitian quantum XY model}
\author{X. Z. Zhang and Z. Song}
\email{songtc@nankai.edu.cn}
\affiliation{School of Physics, Nankai University, Tianjin 300071, China}

\begin{abstract}
We study the geometric phase for the ground state of a generalized
one-dimensional non-Hermitian quantum XY model, which has
transverse-field-dependent intrinsic rotation-time reversal symmetry. Based
on the exact solution, this model is shown to have full real spectrum in
multiple regions for the finite size system. The result indicates that the
phase diagram or exceptional boundary, which separates the unbroken and
broken symmetry regions corresponds to the divergence of the Berry
curvature. The scaling behaviors of the groundstate energy and Berry
curvature are obtained in an analytical manner for a concrete system.
\end{abstract}

\maketitle


\section{Introduction}

\label{sec_intro} Since the existence of an entirely real quantum mechanical
energy spectrum of a non-Hermitian Hamiltonian was proposed in the seminal
work of Bender and Boettcher \cite{Bender 98}, much effort has been devoted
to establish a complex extension of the conventional quantum mechanics \cite%
{Bender 98,Bender 99,Dorey 01,Bender 02,A.M43,A.M,A.M36,Jones}. It has been
shown \cite{AM} that if a Hamiltonian has a symmetry given by an anti-linear
operator $\mathcal{K}$, then either the eigenvalues of the Hamiltonian are
real or they come in complex conjugate pairs. The eigenvector with real
eigenvalue has the symmetry of $\mathcal{K}$, while the one with complex
eigenvalue breaks the symmetry. As system parameter varying, such a sudden
change in the eigenstate can be referred as quantum phase transition in
complex quantum mechanics, while the critical point is called exceptional
point. The characteristic of the critical behavior is the level repulsion,
which leads to the divergence of the first derivative of the eigenvalue with
respect to the system parameter. Although we borrow the concept of QPT from
the conventional Hermitian system, they differ from each other in many
aspects. For instance, the exceptional point can occur in finite
non-Hermitian system, while the QPT is the phenomenon for a Hermitian system
in the thermodynamic limit. This allows the observation of the critical
phenomenon in experiment, since it has been demonstrated that the small size
discrete non-Hermitian system could be realized in optics \cite%
{Bendix,Joglekar,Keya,YDChong,LonghiLaser}.

In this paper, we are interested at the critical behavior of the
eigenfunction in the vicinity of the boundary. In the realm of traditional
quantum mechanics, geometric phase has been introduced to analyze the
quantum phase transitions of the $XY$ model \cite{Carollo,Zhu1,Zhu2}, and
much effort has been devoted to various Hermitian many-body systems \cite{G.
Chen,X. X. Yi,Zanardi,Chenshu,Tao Liu,Ka-Di Zhu,WXG,Fulibin,Ling-Bao Kong}.

We study the geometric phase for the ground state of a generalized
one-dimensional non-Hermitian quantum XY model, which has
transverse-field-dependent intrinsic rotation-time reversal symmetry. Based
on the exact solution, this model is shown to have full real spectrum in
multiple regions for finite size system. The result indicates that the phase
diagram or exceptional boundary, which separates the unbroken and broken
symmetry regions corresponds to the divergence of the Berry curvature.

This paper is organized as follows. In Section \ref{sec_model}, we present
the model Hamiltonian and the solutions. In Section \ref{sec_Phase diagram},
we investigate the phase diagram and analyze the symmetry of the ground
state base on the properties of the solutions. In Section \ref{sec_Geometric
phase}, we give the connection between the phase transition and Berry
curvature. Finally, we give a summary and discussion in Section \ref%
{sec_summary}.

\section{Model and solution}

\label{sec_model}Firstly, we consider a generalized non-Hermitian
one-dimensional spin-$1/2$ $XY$ model in a transverse magnetic field $%
\lambda $ on $N$-site lattice. The Hamiltonian has the form

\begin{eqnarray}
\mathcal{H}\text{ } &\mathcal{=}&\text{ }\frac{J}{4}\sum\limits_{j=1}^{N}%
\left[ \mathcal{G}\sigma _{j}^{+}\sigma _{j+1}^{+}-\mathcal{G}^{\ast
}\left\vert \Lambda \right\vert \sigma _{j}^{-}\sigma _{j+1}^{-}\right.
\label{H_general} \\
&&\left. +\left( \sigma _{j}^{+}\sigma _{j+1}^{-}+\text{H.c.}\right)
+4\lambda \sigma _{j}^{z}\right] ,  \notag
\end{eqnarray}%
where $\mathcal{G}=\mathcal{G}\left( \xi \right) $\ and $\Lambda \left( \xi
\right) $\ are \textit{arbitrary} functions of an $n$-dimensional parameter
vector $\xi =\left\{ \xi _{i}\right\} $\ $i\in \left[ 1,n\right] $, one
component of which is the field strength $\lambda $, i.e. $\xi _{1}=\lambda $%
.\ Here $\sigma _{j}^{\alpha }$ ($\alpha =\pm ,$ $z$) are the Pauli
operators on site $j$, and satisfy the periodic boundary condition $\sigma
_{j}^{\alpha }\equiv \sigma _{j+N}^{\alpha }$. For the sake of simplicity,
we only concern the case of even $N$, the conclusion is available in the
case of odd $N$. We note that the non-Hermiticity of the Hamiltonian arises
from the coupling constants in terms of $\sigma _{j}^{+}\sigma _{j+1}^{+}$
and $\sigma _{j}^{-}\sigma _{j+1}^{-}$. In the case of $\left\vert \Lambda
\right\vert \neq 1$, it represents double spin-flip\ of unequal-amplitude.
In the case of $\mathcal{G}=i\gamma $ (real $\gamma $\textbf{)} and $%
\left\vert \Lambda \right\vert =1$, the Hamiltonian (\ref{H_general})
reduces to
\begin{equation}
\mathcal{H}_{0}\text{ }\mathcal{=}\text{ }J\sum\limits_{j=1}^{N}\left( \frac{%
1+i\gamma }{2}\sigma _{j}^{x}\sigma _{j+1}^{x}+\frac{1-i\gamma }{2}\sigma
_{j}^{y}\sigma _{j+1}^{y}+\lambda \sigma _{j}^{z}\right) ,  \label{H0}
\end{equation}%
which has been investigated in the previous work \cite{zxzspin}. Here we
take the absolute value of the $\Lambda $\ in order to ensure the
non-Hermiticity of the Hamiltonian. Once we relpace $\left\vert \Lambda
\right\vert $\ by $-1$\ in the Hamiltonian (\ref{H_general}), it switches to
a Hermitian operator. Likewise, in this work we define a spin rotation
operator%
\begin{equation}
\mathcal{R}\equiv \exp \left[ -i\phi \sum\nolimits_{j=1}^{N}\sigma _{j}^{z}/2%
\right] ,  \label{R}
\end{equation}%
which has the function of rotating each spin by angle $\phi /2$ about the $z$%
-axis. It turns out that, by taking $\phi =\arg \left( \mathcal{G}\right) $,
we have $\left[ \mathcal{R},\mathcal{H}\right] \neq 0$ and $\left[ \mathcal{T%
},\mathcal{H}\right] \neq 0$, but

\begin{equation}
\left[ \mathcal{RT},\mathcal{H}\right] =0,  \label{RT symmetry}
\end{equation}%
where the antilinear time reversal operator $\mathcal{T}$ has the function $%
\mathcal{T}i\mathcal{T=-}i$, i.e., the Hamiltonian $\mathcal{H}$ is
rotation-time ($\mathcal{RT}$) reversal invariant. In contrast to the
Hamiltonian $\mathcal{H}_{0}$, the rotation is transverse-field-dependent ($%
\lambda $-dependent) through the function $\mathcal{G}$, i.e., $\phi =\phi
\left( \left\{ \xi _{i}\right\} \right) $. It is crucial for the aim of this
paper, revealing the connection between the phase diagram and the Berry
curvature for a non-Hermitian system, that $\phi $\ and $\Lambda $\ are $%
\left\{ \xi _{i}\right\} $-dependent functions. Otherwise, the Berry
curvature vanishes, by no means providing any information of the QPT.

\begin{figure}[tbp]
\includegraphics[ bb=30 280 548 667, width=0.45\textwidth, clip]{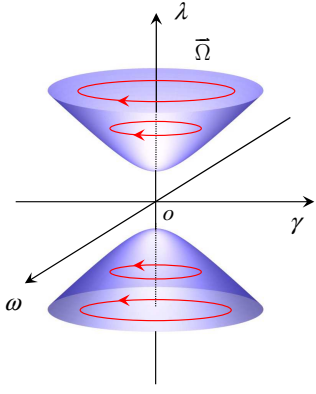}
\caption{(Color online) Schematic of the phase diagram for the Hamiltonian (%
\protect\ref{H_general}) under the condition of Eq. (\protect\ref{example}),
which boundary surface is a hyperboloid of two sheets in $3$-dimensional
space\ $\left\{ \protect\omega ,\protect\gamma ,\protect\lambda \right\} $.
The red circle denotes the force line of the field $\protect\overrightarrow{%
\Omega }$.}
\label{fig1}
\end{figure}


Now we consider the solution of the non-Hermitian Hamiltonian of Eq. (\ref%
{H_general}). We start by taking the Jordan-Wigner transformation \cite{P.
Jordan}%
\begin{eqnarray}
\sigma _{j}^{+} &=&-2\prod\limits_{l<j}\left( 1-2c_{l}^{\dag }c_{l}\right)
c_{j},  \notag \\
\sigma _{j}^{-} &=&-2\prod\limits_{l<j}\left( 1-2c_{l}^{\dag }c_{l}\right)
c_{j}^{\dag }, \\
\sigma _{j}^{z} &=&1-2c_{j}^{\dag }c_{j},  \notag
\end{eqnarray}%
to replace the Pauli operators by the fermionic operators $c_{j}$. Likewise,
the parity of the number of fermions

\begin{equation}
\Pi =\prod_{l=1}^{N}\left( \sigma _{l}^{z}\right) =\left( -1\right) ^{N_{p}}
\end{equation}%
\bigskip is a conservative quantity, i.e., $\left[ \mathcal{H},\Pi \right]
=0 $, where $N_{p}=\sum_{j=1}^{N}c_{j}^{\dag }c_{j}$. Then the Hamiltonian (%
\ref{H_general}) can be rewritten as

\begin{equation}
\mathcal{H}=\sum_{\eta =+,-}P_{\eta }\mathcal{H}_{\eta }P_{\eta },
\end{equation}%
where
\begin{equation}
P_{\eta }=\frac{1}{2}\left( 1+\eta \Pi \right)
\end{equation}%
is the projector on the subspaces with even ($\eta =+$) and odd ($\eta =-$) $%
N_{p}$. The Hamiltonian in each invariant subspaces has the form%
\begin{eqnarray}
\mathcal{H}_{\eta } &=&J\sum\limits_{j=1}^{N-1}\left[ c_{j}^{\dag
}c_{j+1}+c_{j+1}^{\dag }c_{j}-\left\vert \Lambda \right\vert \mathcal{G}%
^{\ast }c_{j}^{\dag }c_{j+1}^{\dag }+\mathcal{G}c_{j+1}c_{j}\right]  \notag
\\
&&-\eta \left[ c_{N}^{\dag }c_{1}+c_{1}^{\dag }c_{N}-\left\vert \Lambda
\right\vert \mathcal{G}^{\ast }c_{N}^{\dag }c_{1}^{\dag }+\mathcal{G}%
c_{1}c_{N}\right]  \notag \\
&&-2J\lambda \sum\limits_{j=1}^{N}c_{j}^{\dag }c_{j}+NJ\lambda .
\label{H_sub}
\end{eqnarray}%
Taking the Fourier transformation

\begin{equation}
c_{j}=\frac{1}{\sqrt{N}}\sum\limits_{k_{\pm }}e^{ik_{\pm }j}c_{k_{\pm }},
\end{equation}%
for the Hamiltonians $\mathcal{H}_{\pm }$, we have

\begin{eqnarray}
\mathcal{H}_{\eta } &=&-J\sum\limits_{k_{\eta }}[2\left( \lambda -\cos
k_{\eta }\right) c_{k_{\eta }}^{\dag }c_{k_{\eta }}  \label{H_eta} \\
&&-\left\vert \mathcal{G}\right\vert \left\vert \Lambda \right\vert
^{1/2}\sin k_{\eta }\left( e^{\beta }c_{-k_{\eta }}c_{k_{\eta }}+e^{-\beta
}c_{-k_{\eta }}^{\dag }c_{k_{\eta }}^{\dag }\right) -\lambda ],  \notag
\end{eqnarray}%
where the momenta $k_{+}=2\left( m+1/2\right) \pi /N$, $k_{-}=2m\pi /N$, $%
m=0,1,2,...,N-1$, and
\begin{equation}
\beta =-\frac{1}{2}\ln \left\vert \Lambda \right\vert +i\left( \phi -\frac{%
\pi }{2}\right) .  \label{beta}
\end{equation}%
Employing the Bogoliubov transformation%
\begin{eqnarray}
A_{k_{\eta }} &=&e^{\beta /2}\cos \left( \frac{\theta }{2}\right) c_{k_{\eta
}}-ie^{-\beta /2}\sin \left( \frac{\theta }{2}\right) c_{-k_{\eta }}^{\dag },
\\
\overline{A}_{k_{\eta }} &=&e^{-\beta /2}\cos \left( \frac{\theta }{2}%
\right) c_{k_{\eta }}^{\dagger }+ie^{\beta /2}\sin \left( \frac{\theta }{2}%
\right) c_{-k_{\eta }},
\end{eqnarray}%
where%
\begin{equation}
\tan \theta =\frac{i\left\vert \mathcal{G}\right\vert \left\vert \Lambda
\right\vert ^{1/2}\sin k_{\eta }}{\left( \lambda -\cos k_{\eta }\right) },
\end{equation}%
one can recast Hamiltonian $\mathcal{H}_{\eta }$ to the diagonal form%
\begin{equation}
\mathcal{H}_{\eta }=\sum\limits_{k_{\eta }}\epsilon _{k_{\eta }}\left(
\overline{A}_{k_{\eta }}A_{k_{\eta }}-\frac{1}{2}\right) ,  \label{H_+/-}
\end{equation}%
with spectrum being%
\begin{equation}
\epsilon _{k_{\eta }}=2J\sqrt{\left( \lambda -\cos k_{\eta }\right)
^{2}-\left( \left\vert \mathcal{G}\right\vert \sqrt{\left\vert \Lambda
\right\vert }\sin k_{\eta }\right) ^{2}}.  \label{spectrum}
\end{equation}

It can be seen that the diagonal form of Hamiltonian $\mathcal{H}_{\eta }$\
in Eq. (\ref{H_+/-}) is still non-Hermitian due to the fact that $\overline{A%
}_{k_{\eta }}\neq A_{k_{\eta }}^{\dag }$. Note that, instead of $\left(
A_{k_{\eta }},A_{k_{\eta }}^{\dag }\right) $, we take the Bogoliubov modes $%
\left( A_{k_{\eta }},\overline{A}_{k_{\eta }}\right) $, which satisfy the
canonical commutation relations%
\begin{eqnarray}
\left\{ A_{k_{\eta }},\overline{A}_{k_{\eta }^{\prime }}\right\} &=&\delta
_{k_{\eta },k_{\eta }^{\prime }},  \label{canon} \\
\left\{ A_{k_{\eta }},A_{k_{\eta }^{\prime }}\right\} &=&\left\{ \overline{A}%
_{k_{\eta }},\overline{A}_{k_{\eta }^{\prime }}\right\} =0.  \notag
\end{eqnarray}%
The eigenstates of $\mathcal{H}_{\eta }$\ and\ $\mathcal{H}_{\eta }^{\dagger
}$\ can be constructed by mode $\left( A_{k_{\eta }},\overline{A}_{k_{\eta
}}\right) $, and the complete biorthogonal bases are established.

In the following analysis, we will focus on the ground state of the
Hamiltonian. It turns out that the ground state lies in the subspace with $%
\eta =+$ in the thermodynamic limit. In addition, the reality of the
groundstate energy can indicate the boundary of the phase diagram. The
ground states of $\mathcal{H}$ and $\mathcal{H}^{\dag }$ are the tensor
product of states in the form
\begin{equation}
\left\vert \text{G}\right\rangle =\prod_{0<k_{+}<\pi }\left[ \cos \left(
\frac{\theta }{2}\right) +ie^{-\beta }\sin \left( \frac{\theta }{2}\right)
c_{k_{+}}^{\dagger }c_{-k_{+}}^{\dagger }\right] \left\vert \text{Vac}%
\right\rangle ,  \label{GS_even}
\end{equation}%
and%
\begin{equation}
\left\vert \overline{\text{G}}\right\rangle =\prod_{0<k_{+}<\pi }\left[ \cos
\left( \frac{\theta }{2}\right) -ie^{\beta ^{\ast }}\sin \left( \frac{\theta
}{2}\right) c_{k_{+}}^{\dagger }c_{-k_{+}}^{\dagger }\right] \left\vert
\text{Vac}\right\rangle ,
\end{equation}%
respectively, with the eigenvalue%
\begin{equation}
E_{g}=-\frac{1}{2}\sum\limits_{k_{+}}\epsilon _{k_{+}}.  \label{E_g}
\end{equation}%
Here $\left\vert \text{Vac}\right\rangle $\ is the vacuum state of the
fermion $c_{j}$.

\section{Phase diagram}

\label{sec_Phase diagram}In this section, we will investigate the phase
diagram of the Hamiltonian (\ref{H_general}) based on the solutions. In a
non-Hermitian system, the term phase diagram has a little different meaning
from that in a Hermitian system. Here it represents the region in which the
non-Hermitian Hamiltonian has full real spectrum or not, rather than the
quantum phase transition \cite{S. Sachdev} in the ground state of a
Hermitian system. From Eq. (\ref{spectrum}), it turns out that the
Hamiltonian (\ref{H_general}) lie in the broken symmetry region when at last
one single-particle level becomes the square root of a negative number, an
imaginary number.


\begin{figure*}[tbp]
\includegraphics[ bb=23 170 550 590, width=0.3\textwidth, clip]{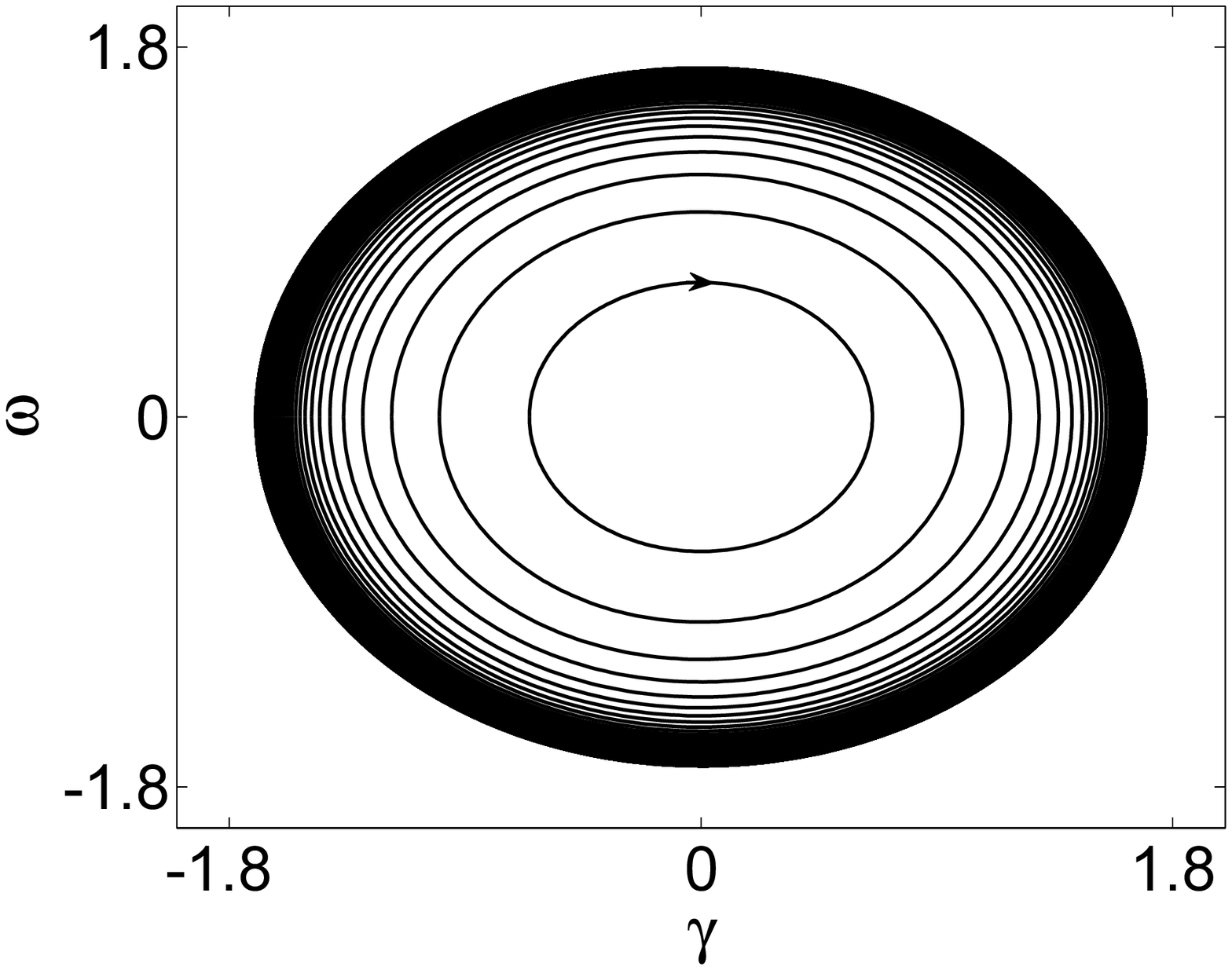} %
\includegraphics[ bb=23 170 550 590, width=0.3\textwidth, clip]{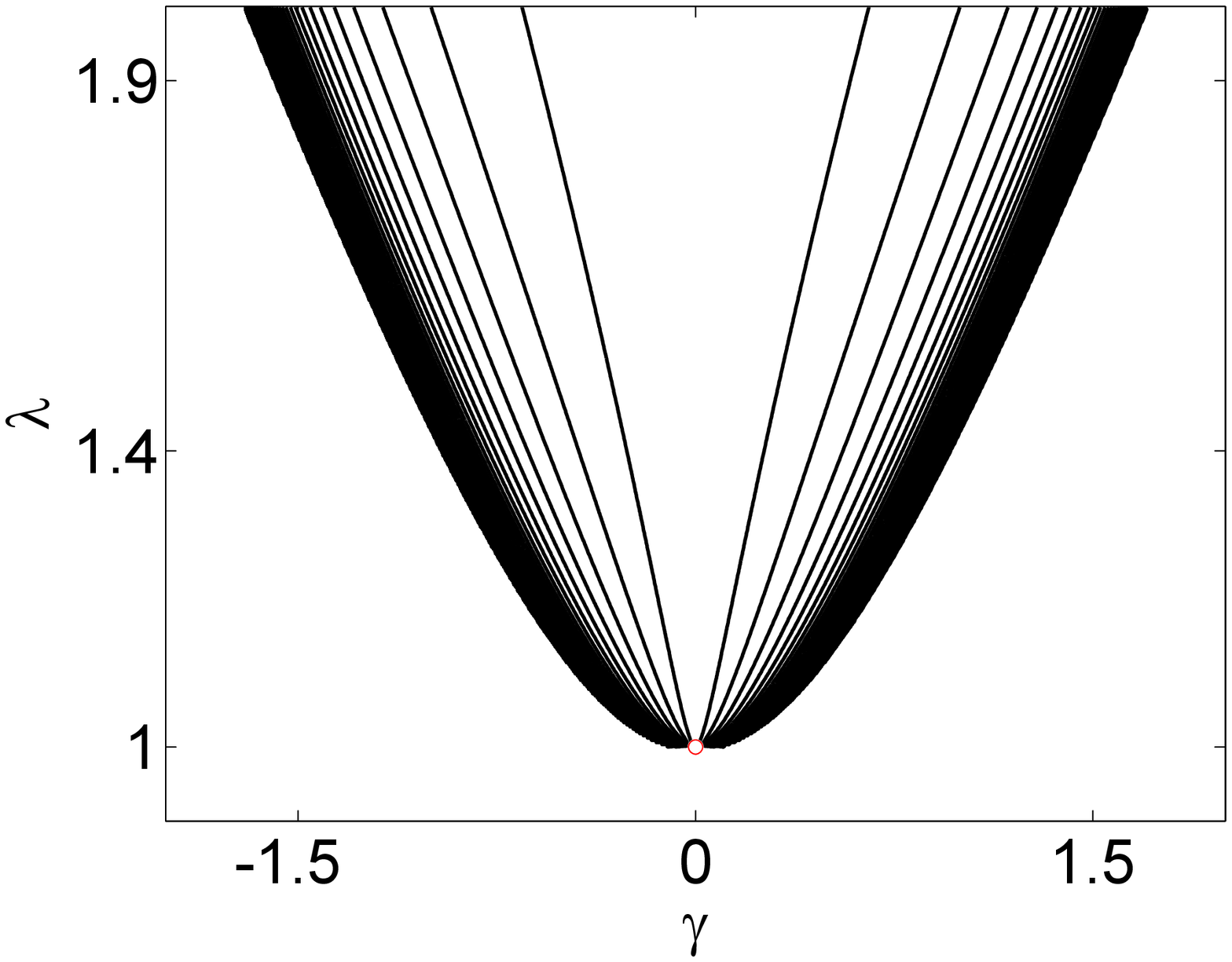}
\caption{(Color online) Contours of the field magnitude $\left\vert \protect%
\overrightarrow{\Omega }\right\vert $\ in (a) $\protect\omega -\protect%
\gamma $ plane with $\protect\lambda =-2$, and (b) $\protect\lambda -\protect%
\gamma $ plane with $\protect\omega =0$, which are obtained from Eq. (%
\protect\ref{magnitude}) for the system with $N=1000$.\quad The contour
interval is constant for the two maps. We see that the pattern of the
contour map accords with our analysis. Note that the broken region does not
include the point $(0,1)$ in $\protect\lambda -\protect\gamma $ plane, which
is denoted by a red circle in Fig. \protect\ref{fig2}(b).}
\label{fig2}
\end{figure*}

Precisely, it is clear that when any one of the momentum $k_{\eta }$\
satisfies%
\begin{equation}
\left\vert \lambda -\cos k_{\eta }\right\vert <\left\vert \mathcal{G}\sin
k_{\eta }\right\vert \left\vert \Lambda \right\vert ^{1/2},
\end{equation}%
the imaginary energy level appears in single-particle spectrum. For finite $%
N $ system, there are totally $2\left( N-1\right) $ equations in the form of
$\left\vert \lambda -\cos k_{\eta }\right\vert =\sqrt{\left\vert \Lambda
\right\vert }\left\vert \mathcal{G}\sin k_{\eta }\right\vert $ for all
possible value of $k_{\eta }$ with$\ k_{\eta }\neq 0,\pi $. It should be
noted that both the $k_{\eta }$\ and $2\pi -k_{\eta }$\ share the same
equations%
\begin{equation}
\lambda =\pm \left\vert \mathcal{G}\right\vert \sqrt{\left\vert \Lambda
\right\vert }\sin k_{\eta }+\cos k_{\eta },
\end{equation}%
which determine the $\left( n-1\right) $-dimensional planes of the fragments
as the boundary of the phase in $n$-dimensional parameter space $\mathcal{V}%
\left\{ \xi _{i}\right\} $ for given functions $\Lambda $\ and $\mathcal{G}$%
. In the thermodynamic limit, the boundary becomes a smooth surface, which
is determined by the following set of parametric equations \cite{Jin Liang}%
\begin{equation}
\frac{\partial \epsilon \left( \xi _{c},k_{\eta }\right) }{\partial k_{\eta }%
}=\epsilon \left( \xi _{c},k_{\eta }\right) =0.
\end{equation}%
Straightforward algebra gives the analytical expression%
\begin{equation}
\lambda _{c}^{2}-\left[ \left\vert \mathcal{G}\left( \xi _{c}\right)
\right\vert \sqrt{\left\vert \Lambda \left( \xi _{c}\right) \right\vert }%
\right] ^{2}=1,  \label{hyperbola}
\end{equation}%
which is the common boundary for both $H_{+}$ and $H_{-}$. Note that the
broken region does not include the surface $\left\vert \mathcal{G}\left( \xi
_{c}\right) \right\vert \sqrt{\left\vert \Lambda \left( \xi _{c}\right)
\right\vert }=0$. In order to illustrate this point, we consider an examples
with $3$-dimensional parameter vectors $\left\{ \omega ,\gamma ,\lambda
\right\} $, and taking
\begin{equation}
\mathcal{G}\sqrt{\left\vert \Lambda \right\vert }=\sqrt{\omega ^{2}+\gamma
^{2}}e^{i\lambda }.  \label{example}
\end{equation}%
The boundary of this examples reads
\begin{equation}
\lambda ^{2}-\omega ^{2}-\gamma ^{2}=1,  \label{boundary}
\end{equation}%
which is a hyperboloid of two sheets in $3$-dimensional space\textbf{\ }$%
\left\{ \omega ,\gamma ,\lambda \right\} $ as shown in Fig. \ref{fig1}.

According to the non-Hermitian quantum theory, the occurrence of the
exceptional point always accomplishes the $\mathcal{RT}$\ symmetry breaking
of an eigenstate. Taking the combination of the Jordan-Wigner and Fourier
transformations on the rotational operator in Eq. (\ref{R}), we have

\begin{equation}
\mathcal{R}\mathcal{=}\prod_{k_{+}}\left[ 1+e^{i\phi }n_{k_{+}}-n_{k_{+}}%
\right] ,
\end{equation}%
where $n_{k_{+}}=c_{k_{+}}^{\dag }c_{k_{+}}$ is the particle number in $%
k_{+} $ space. Applying the $\mathcal{RT}$\ operator on the fermion
operators and its vacuum state $\left\vert \text{Vac}\right\rangle $, we have%
\begin{eqnarray}
\mathcal{RT}c_{k_{+}}^{\dagger }\left( \mathcal{RT}\right) ^{-1} &=&e^{i\phi
}c_{-k_{+}}^{\dagger }, \\
\mathcal{RT}c_{k_{+}}\left( \mathcal{RT}\right) ^{-1} &=&e^{-i\phi
}c_{-k_{+}},  \notag
\end{eqnarray}%
and

\begin{equation}
\mathcal{RT}\left\vert \text{Vac}\right\rangle =\left\vert \text{Vac}%
\right\rangle ,
\end{equation}%
which are available in the both regions. However, the coefficients $\cos
\left( \theta /2\right) $ and $\sin \left( \theta /2\right) $\ experience a
transition as following when the corresponding single-particle level changes
from real to imaginary: We have $\left[ \cos \left( \theta /2\right) \right]
^{\ast }$ $=\cos \left( \theta /2\right) $ and $\left[ \sin \left( \theta
/2\right) \right] ^{\ast }$ $=-\sin \left( \theta /2\right) $ for real
levels and $\left[ \cos \left( \theta /2\right) \right] ^{\ast }$ $=\sin
\left( \theta /2\right) $ for the imaginary levels, respectively. This leads
to the conclusion that the ground state is not $\mathcal{RT}$ symmetric in
the broken region, i.e., $\mathcal{RT}\left\vert \text{G}\right\rangle \neq
\left\vert \text{G}\right\rangle $. It shows that the symmetry of the ground
state can be an indicator of the phase transition as observed in the quantum
phase transition of the Hermitian system. In the following section, we will
investigate the connection between the geometric phase and the phase diagram
in this non-Hermitian spin model, which has been well established for its
Hermitian version \cite{Carollo,Zhu1,Zhu2}.

\section{Geometric phase}

\label{sec_Geometric phase}The Berry curvature for the ground state is an
anti-symmetric second-rank tensor derived from the Berry connection via%
\begin{equation}
\Omega _{ij}=\frac{\partial }{\partial \xi _{i}}\mathcal{A}_{j}-\frac{%
\partial }{\partial \xi _{j}}\mathcal{A}_{i},
\end{equation}%
where
\begin{equation}
\mathcal{A}_{j}=i\left\langle \text{G}\right\vert \frac{\partial }{\partial
\xi _{j}}\left\vert \text{G}\right\rangle _{b}
\end{equation}%
is known as the Berry connection, an $n$-dimensional parameter vector $%
\mathcal{A}=\left\{ \mathcal{A}_{i}\right\} $ in the parameter space. Here $%
\left\langle \ldots \right\vert \ldots \rangle _{b}$ represents the
biorthogonal inner product. Within the unbroken region, we have
\begin{equation}
\mathcal{A}_{i}=-i\frac{\partial \beta }{\partial \xi _{i}}\sum_{0<k_{+}<\pi
}\frac{\left( 1-\cos \theta \right) }{2}.
\end{equation}%
The Berry curvature can be written as
\begin{equation}
\Omega _{ij}=\frac{i}{2}\sum_{0<k_{+}<\pi }\left( \frac{\partial \beta }{%
\partial \xi _{j}}\frac{\partial \cos \theta }{\partial \xi _{i}}-\frac{%
\partial \beta }{\partial \xi _{i}}\frac{\partial \cos \theta }{\partial \xi
_{j}}\right) .
\end{equation}%
Turning back to the $3$-dimensional example system defined in Eq. (\ref%
{example}), a straight forward algebra shows that\textbf{\ }%
\begin{equation}
\overrightarrow{\mathcal{A}}=\mathcal{A}_{\lambda }\widehat{k}
\end{equation}%
where

\begin{equation}
\mathcal{A}_{\lambda }=\sum_{0<k_{+}<\pi }\left[ \frac{1}{2}-\frac{J\left(
\lambda -\cos k_{+}\right) }{\epsilon _{k_{+}}}\right] .
\end{equation}

According to the symmetry of the phase diagram, it is convenient to express
the Berry curvature in cylindrical coordinate system as the form of $%
\overrightarrow{\Omega }=$ $\Omega _{\rho }\widehat{\rho }+\Omega _{\varphi }%
\widehat{\varphi }+\Omega _{\lambda }\widehat{k}$. Here $\widehat{\rho }$, $%
\widehat{\varphi }$ and $\widehat{k}$ denote the unit vectors in cylindrical
coordinate system and $\lambda $ is the axis of symmetry, $\rho =\sqrt{%
\omega ^{2}+\gamma ^{2}}$ is the distance from $\lambda $ axis. The
components are explicitly obtained as

\begin{eqnarray}
\Omega _{\varphi } &=&4J^{3}\rho \sum_{0<k_{+}<\pi }\frac{\left( \lambda
-\cos k_{+}\right) \sin ^{2}k_{+}}{\epsilon _{k_{+}}^{3}},  \label{magnitude}
\\
\Omega _{\lambda } &=&\Omega _{\rho }=0,
\end{eqnarray}%
which shows that $\Omega _{\varphi }$\ is the sum of a series of $\epsilon
_{k_{+}}^{-3}$. It leads to the following features of the curvature: i) The
reality of $\Omega _{\varphi }$, depending on the single-particle spectrum $%
\epsilon _{k_{+}}$,\ is the same as that of the groundstate energy $E_{g}$.
ii) $\Omega _{\varphi }$\ is divergent at every boundary point, the surface
of $\lambda ^{2}-\rho ^{2}=1$. iii) Within the exact symmetric region, the
direction of the field $\overrightarrow{\Omega }$\ is tangent to a circle on
a radius $\rho $ from $\lambda $ axis. In Fig. \ref{fig2} the contours of
the field magnitude $\left\vert \overrightarrow{\Omega }\right\vert $\
obtained from Eq. (\ref{magnitude})\ are plotted schematically. These
features indicate the connection between the geometric phase and the phase
boundary in a pseudo-Hermitian system.

To further understand the relation between geometric phase and quantum
criticality, one can investigate the scaling behavior of the geometric
phases by direct analytical calculation. We start with the investigation of
the scaling behavior for the traditional physical quantity, the groundstate
energy $E_{g}$. According to the axial symmetry of the model, we concentrate
on its first derivatives in $\rho $ and $\lambda $. For the special case in
Eq. (\ref{example}), we have%
\begin{eqnarray}
\frac{\partial E_{g}}{\partial \lambda } &=&-4J^{2}\sum_{0<k_{+}<\pi }\frac{%
\lambda -\cos k_{+}}{\epsilon _{k_{+}}}, \\
\frac{\partial E_{g}}{\partial \rho } &=&4J^{2}\sum_{0<k_{+}<\pi }\frac{\rho
\sin ^{2}k_{+}}{\epsilon _{k_{+}}}.
\end{eqnarray}%
In order to investigate the groundstate energy and Berry curvature
quantitatively and relate their behavior to the criticality, we consider the
two types of approaching paths, parameters tending to the critical point
along the two lines: I) $\left\vert \lambda \right\vert \rightarrow
\left\vert \lambda _{c}\right\vert =\sqrt{\rho ^{2}+1}$ for fixed $\rho $,
and\ II) $\left\vert \rho \right\vert \rightarrow \left\vert \rho
_{c}\right\vert =\sqrt{\lambda ^{2}-1}$ for fixed $\lambda $,\ respectively.
In the thermodynamic limit, we have%
\begin{eqnarray}
\frac{\partial E_{g}}{\partial \lambda } &\rightarrow &-\frac{\sqrt{2}%
J\left( \lambda _{c}^{2}-1\right) ^{1/2}}{\lambda _{c}^{1/2}}\left( \lambda
-\lambda _{c}\right) ^{-1/2}\text{ (I),} \\
\frac{\partial E_{g}}{\partial \rho } &\rightarrow &\frac{\sqrt{2}J\rho
_{c}^{3/2}}{\left( \rho _{c}^{2}+1\right) ^{1/2}}\left( \rho -\rho
_{c}\right) ^{-1/2}\text{ (II),}
\end{eqnarray}%
in the vicinity of the boundary surface.

Now we turn to the investigation of the scaling behavior for the geometric
quantity, the groundstate Berry curvature $\Omega _{\varphi }$. For the
special case in Eq. (\ref{example}), we have%
\begin{equation}
\Omega _{\varphi }=-\frac{1}{4J}\frac{\partial ^{2}E_{g}}{\partial \rho
\partial \lambda }=4J^{3}\rho \sum_{0<k_{+}<\pi }\frac{\left( \lambda -\cos
k_{+}\right) \sin ^{2}k_{+}}{\epsilon _{k_{+}}^{3}}.  \label{relation}
\end{equation}%
Similarly, in the vicinity of the boundary surface we have

\begin{eqnarray}
\Omega _{\varphi } &\rightarrow &\frac{\sqrt{2}\left( \lambda
_{c}^{2}-1\right) }{8\lambda _{c}^{3/2}}\left( \lambda -\lambda _{c}\right)
^{-3/2}\text{ (I),} \\
\Omega _{\varphi } &\rightarrow &\frac{\left( 2\rho _{c}\right) ^{1/2}}{8}%
\left( \rho -\rho _{c}\right) ^{-3/2}\text{ (II).}
\end{eqnarray}%
It indicates that two different approach paths share the same scaling
exponent. Before ending this paper, we want to stress that there is an
interesting relation between the two quantities $\Omega _{\varphi }$\ and $%
E_{g}$, emerging in Eq. (\ref{relation}). It reveals the physical meaning of
the geometrical quantity $\Omega _{\varphi }$\ in this concrete example.
Further work should be done in considering such a relation in the generic
system. Similar work has been done in Hermitian systems, relating the
geometric phase to the energy gap \cite{H. T. Cui}.

\section{Summary}

\label{sec_summary}In summary, we have studied the connection between the
geometric phase and the phase diagram for the pseudo-Hermitian system. We
focused on the ground state of a generalized one-dimensional non-Hermitian
quantum XY model, which has transverse-field-dependent intrinsic
rotation-time reversal symmetry. Based on the exact solution, this model is
shown to have full real spectrum in multiple regions for finite size system.
The result indicates that the phase diagram, which separates the unbroken
and broken symmetry regions corresponds to the divergence of the Berry
curvature. The scaling behavior of the groundstate energy and Berry
curvature are also revealed by the analytical analysis for a concrete
system. The result for such a non-Hermitian quantum spin model may have
profound theoretical and methodological implications.

\acknowledgments We acknowledge the support of the National Basic Research
Program (973 Program) of China under Grant No. 2012CB921900.

\end{document}